\documentclass[fleqn,usenatbib]{rasti}

\usepackage{newtxtext,newtxmath}

\usepackage[T1]{fontenc}

\DeclareRobustCommand{\VAN}[3]{#2}
\let\VANthebibliography\thebibliography
\def\thebibliography{\DeclareRobustCommand{\VAN}[3]{##3}\VANthebibliography}

\usepackage{graphicx}	
\usepackage{amsmath}
\usepackage{siunitx}
\usepackage{courier}

\usepackage{microtype}
\usepackage{graphicx}
\usepackage{booktabs} 

\usepackage{amsmath,amsfonts,bm}

\def\eqref#1{equation~\ref{#1}}

\def\1{\bm{1}}

\def\vv{{\bm{v}}}

\def\vx{{\bm{x}}}
\def\vy{{\bm{y}}}
\def\vz{{\bm{z}}}

\def\mX{{\bm{X}}}

\DeclareMathAlphabet{\mathsfit}{\encodingdefault}{\sfdefault}{m}{sl}
\SetMathAlphabet{\mathsfit}{bold}{\encodingdefault}{\sfdefault}{bx}{n}

\usepackage{subcaption}
\usepackage{caption}

\title[Foundation models for radio astronomy]{Radio Galaxy Zoo: Towards building the first multi-purpose foundation model for radio astronomy with self-supervised learning.}

\author[Slijepcevic et~al.]{
Inigo V.~Slijepcevic$^{1}$\thanks{E-mail: inigo.slijepcevic@postgrad.manchester.ac.uk},
Anna M.~M.~Scaife,$^{1, 2}$
Mike Walmsley,$^{1}$
Micah Bowles,$^{1}$
O.~Ivy Wong,$^{3,4,5}$
\newauthor
Stanislav S.~Shabala,$^{6,5}$
and
Sarah V.~White, $^{7}$
\\
$^{1}$ Department of Physics and Astronomy, University of Manchester, Manchester, UK\\
$^{2}$ The Alan Turing Institute, Euston Road, London, NW1 2DB, UK\\
$^{3}$ CSIRO Space \& Astronomy, PO Box 1130, Bentley, WA 6102, Australia \\
$^{4}$ ICRAR-M468, University of Western Australia, Crawley, WA 6009, Australia \\
$^{5}$ ARC Centre of Excellence for All Sky Astrophysics in 3 Dimensions (ASTRO 3D), Australia \\
$^{6}$ School of Natural Sciences, Private Bag 37, University of Tasmania, Hobart, TAS 7001, Australia \\
$^{7}$ Department of Physics and Electronics, Rhodes University, PO Box 94, Grahamstown, 6140, South Africa
}

\date{Accepted XXX. Received YYY; in original form ZZZ}

\pubyear{2022}

\begin{document}
\label{firstpage}
\pagerange{\pageref{firstpage}--\pageref{lastpage}}
\maketitle

\begin{abstract}
In this work, we apply self-supervised learning with instance differentiation to learn a robust, multi-purpose representation for image analysis of resolved extragalactic continuum images. We train a multi-use model which compresses our unlabelled data into a structured, low dimensional representation which can be used for a variety of downstream tasks (e.g. classification, similarity search). We exceed baseline supervised Fanaroff-Riley classification performance by a statistically significant margin, with our model reducing the test set error by up to half. Our model is also able to maintain high classification accuracy with very few labels, with only $7.79\%$ error when only using 145 labels. We further demonstrate that by using our foundation model, users can efficiently trade off compute, human labelling cost and test set accuracy according to their respective budgets, allowing for efficient classification in a wide variety of scenarios. We highlight the generalizability of our model by showing that it enables accurate classification in a label scarce regime with data from the new MIGHTEE survey without any hyper-parameter tuning, where it improves upon the baseline by $\sim8\%$ . Visualizations of our labelled and un-labelled data show that our model's representation space is structured with respect to physical properties of the sources, such as angular source extent. We show that the learned representation is scientifically useful even if no labels are available by performing a similarity search, finding hybrid sources in the RGZ DR1 data-set without any labels. We show that good augmentation design and hyper-parameter choice can help achieve peak performance, while emphasising that optimal hyper-parameters are not required to obtain benefits from self-supervised pre-training.

\end{abstract}

\begin{keywords}
methods: data analysis -- radio continuum: galaxies -- methods: statistical
\end{keywords}


\section{Introduction}
\label{sec:intro}
\subsection{Current approaches}
To date, supervised CNNs have been the dominant paradigm in radio galaxy morphology classification since their original application in this domain by \citet{Aniyan2017}. The canonical morphological distinction of Fanaroff-Riley I (FRI) and FRII categories has persisted as the most common classification scheme for training and evaluating models - over 40 years since its inception \citep{Fanaroff1974}. However, as sensitivity and resolution has improved, the richness of images has increased leading to a less clear cut and simple picture than suggested by simple binary FR classification \citep{Fanaroff2021AMeerKAT}. This presents great opportunity for more nuanced analysis \citep{Morganti2021CombiningGalaxies} due to the greater quantity of morphological information available from the data, but also requires us to build automated classification tools that are more fine-grained than the FR dichotomy \citep{Mingo2019RevisitingLoTSS}. 

Improvements in efficiency and accuracy have often followed from improvements in architecture within the supervised learning paradigm \citep{Lukic2018, Becker2021CNNClassification, Bowles2021, Scaife2021} and although there is still scope to improve model performance in this way, the gains are diminishing as there is a hard limit set by the amount of information contained in the labelled data. With progress seeming to plateau, it is worth asking the question: given the current available data, is the radio galaxy classification problem largely solved and should we spend more time addressing other aspects of our modelling - for example \citet{Mohan2022QuantifyingClassification} focuses on uncertainty quantification of model predictions rather than raw performance - and/or labelling more data?

Although current supervised models perform well on the straightforward classification task of differentiating FRI/FRII sources, there are a number of weaknesses which can be improved upon:
\begin{itemize}
\item \textbf{Selection bias}: choosing data for labelling using observational factors such as brightness and distance \citep{Hardcastle2020RadioJets} inevitably introduces selection biases.

\item \textbf{Generalization}: supervised learning in astronomy relies on a model trained on a small sub-set of available data to generalize well to a very large possible set of unseen data-points. Our training data has only partial coverage of the full data manifold and therefore a supervised model will inevitably struggle to generalize to the unseen sub-manifolds in observational data.

\item \textbf{Costly labelling}: label cost is high yet many labels are required to train effective models and leverage large models. This problem will be exacerbated with new, high sensitivity telescopes such as the SKA \citep{Dewdney2009TheArray}, where we will have many more resolved sources. 
 
\item \textbf{Implicit bias}: data from new surveys will inevitably challenge the current classification schemes and sources may be mis-classified if we are stuck with models trained using current classification paradigms.

\item \textbf{Computational inefficiency}: models often need to be retrained from scratch when changing/adding/removing classes, training on data from different telescopes or just adding new data. This is an inefficient use of data and computational resources.

\end{itemize}

Given the weakly supervised regime of our data in radio astronomy, i.e. $N_{\rm unlabelled} >> N_{\rm labelled}$, we propose that learning with the much larger quantity of available \emph{unlabelled} data can begin to solve some of these shortcomings in current models, as follows:
\begin{itemize}
    \item If we learn from \textit{all of the available data}, then we have minimized selection biases for our training set and addressed the selection bias issue. Our model will have seen a much larger part of the manifold\footnote{The high dimensional surface which spans all possible data-points in a given data-set.} during train time, improving on the generalization of the model. Although this assumes that we have collected a sufficiency of unlabelled data for our unlabelled training data to be representative of the full data manifold, this is a much weaker assumption than assuming full coverage by the labelled data. We note that while this tackles the selection biases in creating a labelled data-set, there are still significant selection biases involved in creating the unlabelled data-set of cut-out sources from full sky images. Training a model end-to-end directly from sky images is certainly a possible avenue for future progress.
    
    \item By learning from unlabelled data as well as labelled data, we can \textit{reduce the labelling cost} associated with training effective models, reducing the impact of costly labelling. 
    
    \item By \emph{pre-training} models without labels, our models have no encoded bias towards a pre-defined classification scheme, unlike in the transfer learning case, addressing the cause of implicit bias in our models. Furthermore, we can use the same pre-trained backbone for fine-tuning to different downstream tasks, which requires significantly less compute than training from scratch. This greatly reduces the computational resources needed to retrain a model for a new task, e.g. to use a different classification scheme. We can also use our learned representation for scientific applications without labels, such as similarity search.
\end{itemize}

The semi-supervised learning literature includes many effective examples of concurrently learning from labelled data, $\mX_l$, and unlabelled data, $\mX_u$, mostly through smoothing the classification outputs with consistency regularization \citep{Sohn2020FixMatch:Confidence, Berthelot2019MixMatch:Learning, Tarvainen2017MeanResults, Sellars2021LaplaceNet:Classification, Pham2021Meta}. However, although this class of algorithms has been shown to perform well on benchmarking data-sets in computer science, these are usually constructed by simply throwing away labels from a curated data-set to generate ``unlabelled'' data, which ensures that $\mX_l$ and $\mX_u$ are identically distributed. Both from a theoretical perspective \citep{vanEngelen2020ALearning} and empirically in the astronomical domain \citep{Ciprijanovic2020DomainMergers, Slijepcevic2021CanClassification, Slijepcevic2022RadioShift}, it has been demonstrated that when this identically distributed regime is broken (i.e. $\mX_l$ and $\mX_u$ are no longer identically distributed), the performance of this family of algorithms degrades significantly. For this reason, they are less suitable for learning from uncurated, unseen astronomical data.

\subsection{Self-supervised learning}
Self-supervised learning attempts to learn a representation, $\vy_\theta$, given data, $\vx_i \in \mX$, by constructing a task for the model to solve which does not require labels, but forces the model to learn a lower dimensional projection of the data which can help solve not-yet-known downstream tasks with less data and compute. Regardless of the algorithm, the key component of this class of models is an encoder, with the rest of the model being discarded at inference: this is analogous to transfer learning from a supervised model (see \citep{Farahani2020ALearning} for a concise summary of transfer learning), but where our initial training does not require any labels. 

Self-supervised representation learning techniques have recently been shown to generate highly generalizable representations with state-of-the-art performance on downstream tasks. Contrastive learning methods such as SimCLR \citep{ChenSIMCLR}, MoCo \citep{He2020MomentumLearning}, and BYOL \citep{Grill} (\citet{Jaiswal2020ALearning} provide a survey) rely on instance differentiation\footnote{A loss constructed by using augmentations to generate stochastically perturbed views of the same image} and work well with the convolutional networks that have become ubiquitous in astronomical image analysis. More recently, the vision transformer architecture \citep{DosovitskiyViT} has enabled effective learning on extremely large data-sets due to its computational efficiency and flexible inductive bias which can be learned during training, with some algorithms using instance differentiation \citep{Caron2021EmergingTransformers} as in the convolutional case, and others applying novel loss functions such as a pixel reconstruction loss \citep{He2022MaskedLearners}. Furthermore, numerous studies have recently shown an improved robustness to distribution shift, class imbalance and out-of-distribution (o.o.d.) data \citep{Liu2021Self-supervisedImbalance, Zhong2022IsLearning, Shi2022HowShift} as well as transferability \citep{MoeinShariatnia2022HowTransfer} of pre-trained models compared with their supervised counterparts. 

Within astronomy, there have been some attempts to use unlabelled data to improve performance in various tasks. \citet{Hayat2020EstimatingLearning} successfully improve upon state of the art galactic distance estimations using images;  \citet{Marianer2021ASources} tackle outlier detection in gravitational wave data; \citet{Richards2012Semi-supervisedClassification} attempt to improve photometric supernova classification. \citet{Spindler2020} leverage variational auto-encoders and Gaussian mixture models to perform both image generation and unsupervised clustering. More recently, \citet[][]{Hayat2021Self-SupervisedImages} and \citet[][]{Stein2021} have inspired our work with their successful applications of contrastive learning methods to galaxy morphology classification and gravitational lens/non-lens classification, respectively; \citet{Walmsley2022TowardsLearning} show performance improvements through combining contrastive learning with a novel supervised loss for vote prediction with a data-set from citizen science; and an initial study in \citet{Slijepcevic2022LearningLearning} showed that these contrastive techniques show great promise for pre-training in radio astronomy.

We also note the radio astronomy community is producing larger, more varied catalogues of extended sources, causing traditional classification schemes to be questioned. \citet{Rudnick2021RadioBoxes} suggests a more flexible approach based on tagging sources rather than classifying. Given that a self-supervised model does not have any implicit bias towards a given classification scheme, it is a naturally more complementary method to tagging than conventional supervised learning, and future work could explore combining these two paradigms.

\subsection{This work}
In this work, we apply Bootstrap Your Own Latent \citep[BYOL;][]{Grill}, a self-supervised learning method based on instance differentiation to our data-set of radio galaxies, comparing its performance to a supervised baseline and discussing the merits of using such an approach.

We are careful to modify the original algorithm as required, specifically with regard to the augmentation pipeline: we present detailed results explaining these design choices in Section~\ref{subsec:augs}. This is important as radio galaxy images are qualitatively different from terrestrial images: classes are generally much more fine-grained, and information is easily destroyed when heavily augmenting images, as well as fewer augmentations being available due to the greyscale nature of the images.

We hope that in further developing self-supervised learning in a domain-specific way for radio astronomy and astronomy more generally, we can begin to develop \textit{foundation models} \citep{Bommasani2021OnModels} that can be applied a wide variety of data analysis tasks in the field.

\vskip .1in
\noindent
The main contributions of this work are:
\begin{itemize}
    \item Demonstrate the improvement in classification accuracy achieved by pre-training with a large unlabelled data-set and fine-tuning, both in the low labelled data regime \textit{and} with all labels available.     

    \item Give evidence that learning from a large unlabelled data-set from a single source produces a model with strong generalization across surveys.
    
    \item Show that our model can perform useful scientific tasks such by finding many hybrid sources in the RGZ DR1 (Radio Galaxy Zoo Data Release 1) data-set using only a single known hybrid source.
    
    \item Quantify the importance of different augmentations when applying the BYOL algorithm to novel data in the radio astronomy domain, finding that although random resized cropping and rotation are key for maximal model performance, even sub-optimal augmentations produce strong results as long as too many are not removed.

    \item Show fine-tuning atleast 1 layer is extremely important for classification accuracy but that that deeper fine-tuning of our model only yields better performance if enough data is available.
    
\end{itemize}

In Section~2.1-2.4 describe the algorithms and data used to train our model and provide training details (e.g. augmentations, warmup scheduler). Section~2.5 covers the . and describe the process of evaluating our trained models. Sections~3.1-4 compare the performance of our model to the baseline and give possible examples of scientific applications. Sections~3.4-3.7 provide experiments justifying choices made when tuning the model. Section~3.8 gives presents the results of ablating out augmentations. Section 4 concludes the paper.
\section{Method}

\label{sec:method}
\subsection{Bootstrap Your Own Latent (BYOL)}
\label{subsec:BYOL}
In the BYOL algorithm \citep{Grill}, the model is trained to minimise the distance between different random augmentations of the same image in the representation space of the encoder. Some contrastive learning algorithms, such as SimCLR \citep{ChenSIMCLR}, also teach the model to maximize the distance to augmentations of other images further away, but recent work has shown that this is not needed to achieve state-of-the-art performance, with BYOL out-performing SimCLR when trained on ImageNet \citep{Deng2009ImageNet:Database}. There are significant benefits to omitting these ``negative pairs'' during training: primarily, significantly less computation time and a much lower memory footprint, which enables faster iteration within a finite compute budget. Furthermore, astronomical images have a fuzzier semantic meaning, with less clear cut differences between classes, increasing the probability of class collision \citep{Arora2019ALearning} when sampling negative pairs. For these reasons, we use BYOL as our core learning algorithm.

BYOL uses a momentum encoder to calculate positive pair losses, which helps to avoid representation collapse, although it has been shown that this is not strictly required \citep{ChenSiamese}. An exponential moving average of the online network parameters, $\theta$, gives the parameters of the momentum encoder, $\xi$, which are updated at each step such that
\begin{equation}
    \label{eq:momenc}
    \xi_i \leftarrow \tau \xi_{i-1} + (1-\tau) \theta,
\end{equation}
where the hyperparameter $\tau$ gives the decay rate.

Both the momentum and online encoder have fully connected projection heads, which output $\vz_\xi$ and $\vz_\theta$, respectively. The online network has an additional fully connected prediction head, $q_\theta$. These extra layers separate the loss from $\vy_\theta$, preventing the model from overfitting to the self-supervised task and generating a representation space that generalizes better to many downstream tasks. Experiments have shown this to be the case empirically when pre-training and fine-tuning to a specific task \citep{ChenSIMCLR}.

\begin{figure*}
\centering
    \centering
    \includegraphics[width=0.9\linewidth]{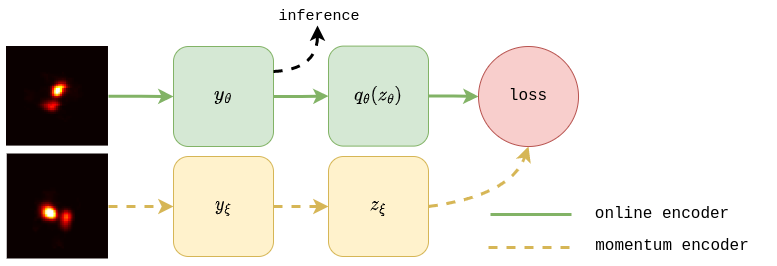}
    \caption{Diagram showing the flow of tensors in the BYOL algorithm. Dashed lines indicate that gradients are \textbf{not} back-propagated}
\label{fig:byol}
\end{figure*}

Two augmentations, $t$ and $t'$, are drawn from the same augmentation pool, $\mathcal{T}$, giving views $\vv$ and $\vv'$ of the original image, $\vx$. $\vv$ [$\vv'$] is passed through the online [momentum] encoder to give $q_\theta (z_\theta)$ [$\vz_\xi$]. The loss 
\begin{equation}
    \label{eq:byol_loss}
    \mathcal{L} = || q_\theta ( \vz_\theta ) - \vz_\xi ||^2_2,
\end{equation}
 is the mean squared error between the representations of the momentum encoder and the online encoder. Only the online network parameters, $\theta$, are updated by back-propagation, with momentum encoder parameters, $\xi$, updated using an exponential moving average as given in Equation~\ref{eq:momenc}. In Figure~\ref{fig:byol}, we illustrate the flow of tensors through the model\footnote{Our implementation is publicly available at \url{https://github.com/inigoval/byol}}. 
 

\subsection{Data} 
\label{subsec:data}
We use the MiraBest data-set \citep{PorterMiraBest2023}, which has 1,256 sources and a roughly even FRI/FRII split as our labelled data for fine-tuning, validating and testing. Each sample in the MiraBest data-set is also tagged with a \textit{Confident} or \textit{Uncertain} tag, which corresponds to the confidence of the human classification of the data point. We omit \textit{Uncertain} samples from the training set in our experiments as we find it makes the evaluation noisier without any significant gain in performance, leaving us with 833 \textit{Confident} images. We do however use these \textit{Uncertain} data for evaluation, as we can use them to test our model's out-of-distribution generalization. 


We use a sub-set of 104 \textit{Confident} samples as our held out test set, which has been chosen to be representative of the full data-set by choosing at least one example of each sub-class within the data-set. This test-set has been used to benchmark the MiraBest data-set in the past and makes comparison with past and future models straightforward. Unless otherwise specified, we use 25\% of the remaining data to validate our model and choose optimal hyper-parameters, and the final 75\% for training. Due to the very small number of labels, we do not use a validation set when reporting test accuracy and use 100\% of the training data for training.

For our unlabelled data-set, we use the RGZ DR1 data-set (Wong et al. 2023 in prep) which contains 107,893 images, has not been curated and may contain out-of-distribution data-points, class imbalance or irregular sub-population distributions. For the RGZ DR1 data-set we also have the angular size in arcseconds for each radio source, which has been calculated algorithmically. Both of these data-sets use images from the VLA FIRST survey \citep{Becker1995TheCentimeters}, and some of the MiraBest samples are contained in the RGZ DR1 data-set - we have removed these from RGZ DR1 to avoid leakage of the validation/test sets into the training set. Further information regarding these data-sets can be found in \citet{Slijepcevic2022RadioShift}.

\subsection{Augmentation scheme}
\label{subsec:intro_augs}
Due to the self-supervised instance differentiation task, the BYOL algorithm relies heavily on an appropriate augmentation scheme. For this reason we customise the standard ImageNet augmentation pipeline \citep{ChenSIMCLR} for our purposes, which we provide a discussion of below. We give the augmentations in the order in which they are performed. We also perform an ablation study to show the importance of different augmentations in Section~\ref{subsec:augs}

\begin{itemize}
    \item \textbf{Rotation}: Given the rotational invariance of astronomical sources, we want to make sure that our model does not structure the representation space with respect to the orientation of the images (i.e. we want a representation that is invariant to the image rotation). For this reason, we rotate every input image to a random orientation with nearest neighbour interpolation to re-grid the image.
    
    \item \textbf{Center crop}: We center crop the $128 \times 128$ images to $70 \times 70$ (details in Section~\ref{subsec:augs}). 
    
    \item \textbf{Random resized crop}: We reduce the range of random cropping to $80-100\%$ of the image (details in Section~\ref{subsec:augs}).
    
    \item \textbf{Random flipping}: We randomly flip every image both horizontally and vertically, to add stochasticity and remove any bias due to chirality in the images.
    
    \item \textbf{Color jitter}: Despite our images having only one color channel, color jitter still changes the image through contrast and saturation. We use color jittering parameters $ P( \texttt{color\_jitter} ) = 0.8 $ and $s = 0.5$.
    
    \item \textbf{Blur}: We apply blurring with a low probability of 0.1 (see Section~\ref{subsec:augs} for details).
    
\end{itemize}

\begin{table}
\begin{center}
\begin{tabular}{cc} 
 Augmentation           & Hyper-parameter value\\ 
 \hline
 Rotation               & 360 degrees (uniform distribution) \\
 Center crop            & 70 pixels \\
 Random resized crop    & 80-100 \% \\
 Horizontal Flip        & $p = 0.5$ \\
 Vertical Flip          & $p = 0.5$ \\
 Color Jitter           & $s = 0.5$, $p=0.8$ \\
 Blur                  & $ p=0.1$ \\
\end{tabular}
\end{center}
\caption{Augmentation pipeline used during self-supervised pre-training.}
\label{tab:augs_byol}
\end{table}

\begin{table}
\begin{center}
\begin{tabular}{cc} 
 Augmentation           & Hyper-parameter value\\ 
 \hline
 Rotation               & 360 degrees (uniform distribution) \\
 Center crop            & 70 pixels \\
 Random resized crop    & 90-100 \% \\
 Horizontal Flip        & $p = 0.5$ \\
Vertical Flip           & $p = 0.5$ \\
\end{tabular}
\end{center}
\caption{Augmentation pipeline used during supervised training and fine-tuning/linear evaluation.}
\label{tab:augs_sup}
\end{table}

During self-supervised training, we use all of the augmentations above, see Table~\ref{tab:augs_byol} for a summary, whereas during fine-tuning/linear probing/supervised baseline training we use only geometric/affine augmentations, see Table~\ref{tab:augs_sup} for details.
 
\subsection{Training details} 
\label{subsec:trainingdetails}
We use the ResNet architecture \citep{He2016DeepRecognition} to train both our supervised baseline and self-supervised models. The ResNet architecture consists of stacking convolutional blocks which have residual skip connections, allowing deeper models to converge and reap the benefits of scaling. ResNets are the standard for comparisons of both supervised and self-supervised algorithms and have repeatedly been shown to be capable of state-of-the-art performance in the 5 years since their inception \citep{Wightman2021ResNetTimm}.

During pre-training of BYOL, we only use the RGZ DR1 data-set, i.e. none of the images from the MiraBest data-set are shown to the model. During fine-tuning and evaluation of our pre-trained model we use \textit{Confident} samples from MiraBest.

We pre-train until convergence of the \textit{training} loss. This can be sub-optimal, and it is likely that we can achieve better performance on FR classification by validating on the labelled data set and using early stopping. However, we wish to emulate the most general setting where we are training a foundation model on purely unlabelled data, and where there are no labels available at all until finetuning, so we simply train until convergence of the similarity loss. Furthermore, by doing this we avoid over-fitting to a validation set, which might bias our results \citep{Wightman2021ResNetTimm}.
When training both BYOL and the supervised baseline we use the same backbone model for feature extraction (e.g. ResNet-18). BYOL uses multiple fully connected heads during training as explained above, although these are thrown away at inference (details can be found in \citealt{Grill}). 

We use a linear warmup cosine scheduler with 10 epochs of warm-up with the Stochastic Gradient Descent (SGD) optimizer. We follow \citet{Grill} for our hyperparameters, which we summarise in Table~\ref{tab:hparams}.

Other than when we are varying the number of available labels (Section~\ref{subsec:classification}), we use 75\% of available data for training, and 25\% of available data for validation.

\begin{table}
\begin{center}
\begin{tabular}{cc} 
 Hyperparameter     & Value\\ 
 \hline
Optimizer           & SGD \\
Learning Rate       & $0.2$ \\ 
Batch Size          & 1024 \\ 
Weight Decay        &  \num{1.5e-6} \\ 
Scheduler           & Linear warmup cosine decay \\ 
Warmup epochs       & 10 \\
Epochs              & 500 \\
\end{tabular}
\end{center}
\caption{Hyperparameters used for BYOL training. The learning rate given is the base learning rate, which is scaled by the batch size by a factor $\frac{batch \: size}{256}$.}
\label{tab:hparams}
\end{table}

\subsection{Supervised Baseline}
\label{subsec:baseline}

\begin{table}
\begin{center}
\begin{tabular}{cc} 
 Hyperparameter     & Value\\ 
 \hline
Optimizer           & AdamW \\
Base Learning Rate  & $0.001$ \\ 
Batch Size          & 64 \\ 
Weight Decay        &  \num{0.005} \\ 
Beta 1              &  \num{0.9} \\ 
Beta 2              &  \num{0.999} \\ 
Scheduler           & Cosine decay \\ 
Epochs              & 300 \\
Layer-wise decay  (fine-tuning only)  & 0.75 
\end{tabular}
\end{center}
\caption{Hyperparameters used for finetuning the model (training the supervised baseline). The learning rate given is the base learning rate, which is scaled by the batch size by a factor $\frac{batch \: size}{256}$.}
\label{tab:hparams_finetune}
\end{table}

We need to ensure that we do not bias our results in favour of a ``new'' technique or it can be difficult to disambiguate the factors which lead to an improvement in performance \citep{Wightman2021ResNetTimm}. We therefore use the same ResNet architecture for our supervised baseline as for pre-training and the same linear classification head used when evaluating our pre-trained model. The supervised baseline is trained on \textit{Confident} MiraBest only, and evaluated on the held out \textit{Confident} test samples, as described in Section~\ref{subsec:data}. Due to the different nature of training in the supervised case, we use a much smaller batch size and fewer epochs. During fine-tuning of our pre-trained model we use the same hyper-parameters as the supervised baseline. We detail the hyperparameter setup in Table~\ref{tab:hparams_finetune} and the augmentation pipeline in Table~\ref{tab:augs_sup}.

Our baseline implementation is fully reproducible using our public GitHub code\footnote{\url{https://github.com/inigoval/supervised}}.

\subsection{Evaluation}
\subsubsection{Linear evaluation}

\begin{table}
\begin{center}
\begin{tabular}{cc} 
 Hyperparameter     & Value\\ 
 \hline
Optimizer           & SGD \\
Base Learning Rate  & $0.1$ \\ 
Batch Size          & 64 \\
Momentum            & 0.9 \\
Weight Decay        & 0 \\ 
Scheduler           & None \\ 
Warmup epochs       & 10 \\
Epochs              & 100
\end{tabular}
\end{center}
\caption{Hyperparameters used for linear probing the supervised baseline. The learning rate given is the base learning rate, which is scaled by the batch size by a factor $\frac{batch \: size}{256}$.}
\label{tab:hparams_linprobe}
\end{table}

To measure the quality of our learned representation, we use the linear evaluation protocol as commonly used when evaluating self-supervised models \citep{ChenSIMCLR}. Linear evaluation is performed by training a single linear layer (i.e. a classical logistic regression model) on top of the representation with labels for the given task, and evaluating the accuracy on a held out test set. The underlying pre-trained model wegihts are frozen and not updated. 

The linear evaluation protocol is not meant to represent the final accuracy of the model for a downstream task. The goal of linear evaluation is to test how well separated the representation space learned by the self-supervised model is for a specific downstream task, such as FRI/FRII classification, \textit{without the model learning anything about this specific classification scheme explicitly}. This evaluation technique asks the question: "How useful are the weights obtained by training our model on unlabelled data with a self-supervised objective for a specific classification question (FR classification in our case)". Linear evaluation also has the advantage of being computationally extremely cheap, which means we can perform this evaluation during training for our self-supervised model in order to monitor training. 

The hyper-parameters used for linear evaluation are not designed to achieve minimum test set error, rather to transparently evaluate the representation. By fine-tuning the weights of the pre-trained model also, the difference in quality of the pre-trained representations will decrease as the network learns to separate the classes during fine-tuning. We give the full list of hyperparameters used in Table~\ref{tab:hparams_linprobe} and the augmentation pipeline in Table~\ref{tab:augs_sup}

Our linear evaluation implementation can be found on GitHub \footnote{\url{https://github.com/inigoval/byol} \label{foot:finetune}}.

\subsubsection{Fine-tuning}
In practice, when aiming to maximise accuracy on a downstream task, we would also fine-tune deeper layers of the model. We therefore also test our model in this training scheme, by fine-tuning the model to different depths with a learning rate which decays linearly as we go deeper into the model. The learning rate of the i$^{th}$ layer is given by 
$LR_{i} = LR_{base} \times \rho^i$
where $\rho$ is the layerwise decay rate. 

Fine-tuning is done at test time once the model has been trained, as it is too computationally expensive to perform as a validation metric. Unlike the linear evaluation case, with fine-tuning we are attempting to achieve maximum performance. In order to keep the comparison fair, we keep our hyperparameters consistent between the fine-tuning and supervised learning cases. We give the full list of hyperparameters used in Table~\ref{tab:hparams_finetune} and the augmentation pipeline in Table~\ref{tab:augs_sup}.

Our fine-tuning implementation can be found on GitHub\textsuperscript{\ref{foot:finetune}}.

\subsubsection{Reproducibility}

We run all our validation experiments with multiple seeds and average over the results, with a standard error reported for each result. This includes averaging over different data splits for our training and validation set (when fine-tuning/linear probing) and initialization weights for the linear head layer. Furthermore, by not using early stopping but training until the final epoch, we can avoid some stochasticity in our results \citep{Wightman2021ResNetTimm}.

Our test set is held out and only used for final evaluation. The test set is chosen to be representative of the MiraBest data-set not just with respect to FRI/FRII classes but also more fine-grained sub-classes.

\section{Experiments}

\subsection{Dimensionality reduction and visualisation}
\label{subsec:vis}

\begin{figure}
\centering
    \centering
    \includegraphics[width=\columnwidth]{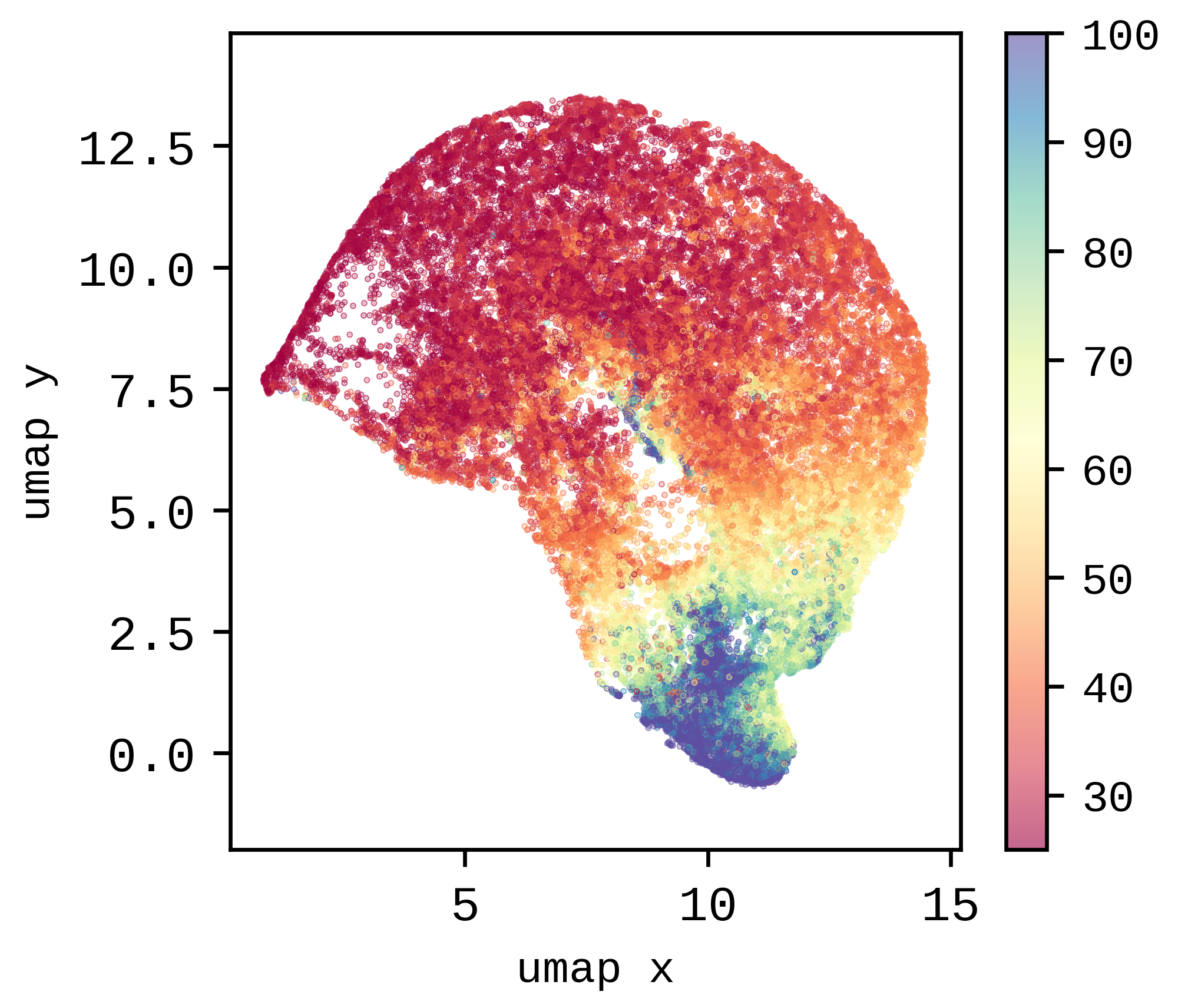}
    \caption{RGZ DR1 data-set visualized using UMAP (after being passed through our pre-trained model first). The color bar shows the source extension of each data point in arcseconds. This visualization can be dynamically explored using our \href{https://github.com/inigoval/rgz-latentexplorer}{webapp}. X and y directions have no explicit physical meaning.}
\label{fig:umap_rgz}
\end{figure}

\begin{figure}
\centering
    \centering
    \includegraphics[width=\columnwidth]{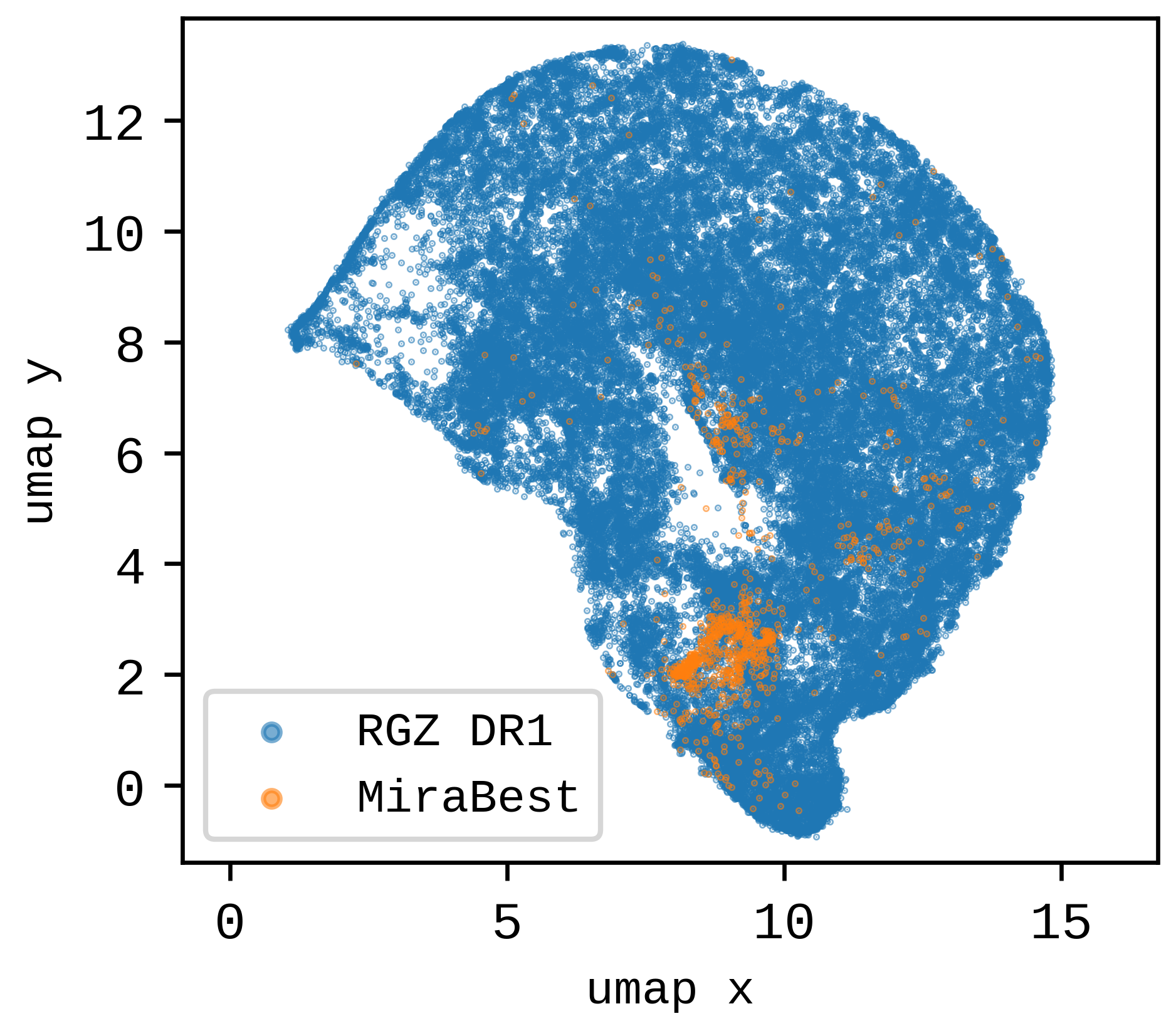}
    \caption{RGZ DR1 and MiraBest data-sets visualized using UMAP after being being encoded through our pre-trained model. Note that there is some visual difference in the representation here as compared with Figure 10 which is due to the addition of the MiraBest data-set when fitting UMAP. X and y directions have no explicit physical meaning.}
    
\label{fig:umap_mbrgz}
\end{figure}

It is often useful to visualize data in a way that humans can view. Although self-supervised learning significantly reduces the dimensionality of the input data (from 4900 to 512), it is still too high dimensional to easily represent the data in a way that humans can parse. To reduce our data down further, to two dimensions which we can easily view, we can pass the data through our self-supervised model, and then reduce the data down further algorithmically. 

There are many ways of reducing the dimensionality of a data-set in this way, such as principal component analysis (PCA), t-SNE \citep{VanDerMaaten2008VisualizingT-SNE} and Uniform Manifold Approximation and Projection \citep[UMAP; ][]{McInnes2018}. To reduce the dimension of the feature space of the encoder to a visualizable 2D space, we use apply a combination fo PCA and UMAP. Since the feature space of our model has some redundancy, we first reduce the dimensionality to 200 with PCA, which preserves 99.7\% of the variance. We compress the dimensionality further to 2 dimensions with UMAP, since it preserves global structure and scales better than t-SNE with dimensionality and data-set size. It is important to understand that the x and y directions in these visualizations are not physically meaningful.

In Figure~\ref{fig:umap_rgz} we visualize the entire RGZ DR1 data-set, showing that our representation is structured with respect to source extension, in spite of the fact that this information was never explicitly shown to the model. This highlights and confirms that by solving the instance differentiation task, our model learns to represent meaningful characteristics of the data.  

In Figure~\ref{fig:umap_mbrgz}, we demonstrate that the MiraBest data-set only occupies a small sub-manifold of the full RGZ data-set (in the representation space of our pre-trained model). This is due to selection biases in the labelling process, and we can see quite clearly that the labelled data-set is biased towards large angular size sources. This means that a model trained only on this data will struggle to make reliable predictions on data with small angular size sources. This highlights the benefits of pre-training, such that the model is exposed to the full data manifold, which should allow more robust inference on a wider range of test data. In Section~\ref{subsec:classification} we confirm that this is beneficial to both in-distribution and out-of-distribution classification tasks.

To make this visualization of the model's representation space available to radio astronomers, we provide an interactive app where the model can be queried with an arbitrary image from the VLA FIRST survey by inputting sky co-ordinates. We encourage anyone interested in our results to use it - installation takes only a few minutes and no machine learning or coding experience is required to use it. The app can be found on GitHub\footnote{\url{https://github.com/inigoval/rgz-latentexplorer}}.

\subsection{Downstream task 1: Fanaroff-Riley classification with limited labels}
\label{subsec:classification}

\begin{table}
\label{tab:conf_acc}
\begin{center}
\begin{tabular}{c|cc} 
\# of labels    &  Supervised Error (\%) & Fine-tuned Error (\%) \\ 
\hline
36              & $19.03 \pm 1.39$          & \bm{$16.15 \pm 1.00$} \\
72              & $18.75 \pm 0.73$          & \bm{$11.83 \pm 0.11$} \\
145             & $11.92 \pm 0.50$          & \bm{$7.79 \pm 0.76$} \\ 
291             & $7.02 \pm 0.63$          & \bm{$4.90 \pm 0.53$} \\ 
437             & $5.77 \pm 0.54$          & \bm{$4.52 \pm 0.52$} \\
583             & $5.10 \pm 0.29$          & \bm{$4.33 \pm 0.36$} \\
656             & $4.81 \pm 0.16$          & \bm{$3.08 \pm 0.35$} \\
729             & $4.81 \pm 0.16$          & \bm{$1.92 \pm 0.25$} \\
\end{tabular}
\caption{MiraBest (\textit{Confident}) test set error when trained on the full \textit{Confident} subset (729 samples). All values are averaged over 10 runs with different seeds for model parameter initialization.}
\end{center}
\end{table}

\begin{figure}
\centering
    \centering
    \includegraphics[width=\columnwidth]{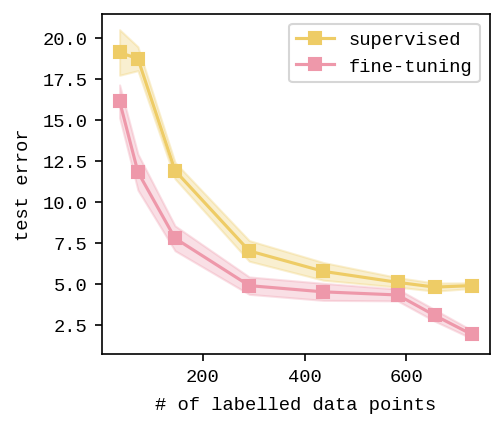}
    \caption{Test set error on the MiraBest \textit{Confident} subset as a function of number of labels. In the BYOL case, we fine-tune the final 10 layers. When using the smallest number of labels (26), we only train for 100 epochs. Error bars show the standard error.}
\label{fig:main_acc}
\end{figure}

An important part of the benefits of pre-training is that labels can be used more efficiently than when training from scratch. During the pre-training stage, the model learns a representation from all available unlabelled data, which allows label efficient training for a specific downstream task (e.g. FRI/II classification). 


In Figure~\ref{fig:main_acc}, we see that the pre-trained and fine-tuned model outperforms the supervised baseline by a significant margin regardless of the number of labels available. This is in contrast to some previous studies, which only show improvements in a specific label regime \citep{Slijepcevic2022RadioShift}. Given that the self-supervised model performs better even with all labels available, we can infer that our labelled data-sets are not large enough to learn a model which generalizes well enough to achieve maximal performance on the test set classification task. We hypothesise that by pre-training we are able to leverage model scaling as deeper layers in the network have already been trained with a larger volume of unlabelled data, reducing the amount of overfitting when training with the labelled data. It highlights that, during training, it is beneficial for the model to traverse a larger portion of the data manifold (when pre-training), even when the test set is drawn only from a small sub-manifold.

At low label volumes, the advantage of pre-training is even more pronounced. We see that with only 72 labels, the pre-trained model can still achieve $\sim 88 \%$ accuracy, whereas our supervised model performs $\sim 9 \% $ worse. With 291 labels, the pre-trained model achieves the same accuracy as the supervised model with all 729 available labels. In the context of training a model on new scientific data, our foundation model therefore reduces the volume of labelled data required by a factor of almost 3. Reducing the labelling cost of training an effective model is extremely advantageous as it both reduces both the training time \textit{and} the human cost of labelling, allowing us to iterate and test new models and classification schemes faster and with less cost.

\begin{figure}
\centering
    \centering
    \includegraphics[width=\columnwidth]{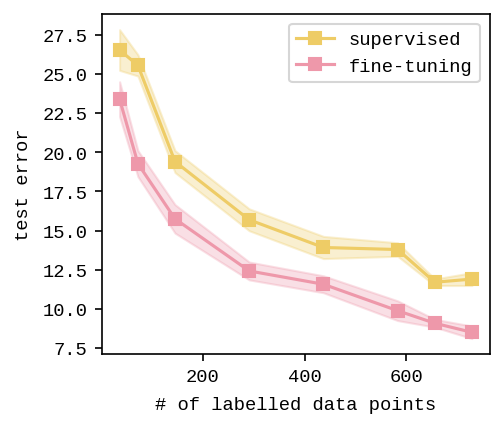}
    \caption{Test set error on the MiraBest \textit{Uncertain} subset as a function of number of labels. In the BYOL case, we fine-tune the final 10 layers. When using the smallest number of labels (26), we only train for 100 epochs. Error bars show the standard error.}
\label{fig:unc_acc}
\end{figure}

The MiraBest data-set is only a small sub-manifold of the full available RGZ DR1 data-set (this is demonstrated in Section~\ref{subsec:vis}). We test whether there is measurable benefit to the model seeing the whole data manifold during pre-training rather than only being exposed to the small labelled sub-manifold as is the case with purely supervised training. Specifically, we hypothesise that the model may be able to deal with out-of-distribution data that is not in the labelled data, but which the model may have seen during pre-training. We can extend our previous experiment quite naturally to this case, by taking the pre-trained model fine-tuned only on \textit{Confident} samples and evaluating it at test time on \textit{Uncertain} samples from the MiraBest data-set. In Figure~\ref{fig:unc_acc} we can see that our fine-tuned model significantly outperforms the baseline at all label volumes - even more so than in the in-distribution case previously discussed. This is not surprising: the baseline is trying to generalize to a sub-population of data it has not seen, whereas the pre-trained model has learned useful features about the \textit{Uncertain} part of the data manifold which should lead to more robust predictions. Furthermore, we note that the fine-tuned trend in Figure~\ref{fig:unc_acc} suggests that the model would be able to further improve its predictions given more data, which is not the case for the baseline. This highlights the ability of a fine-tuned model to more efficiently utilise a large capacity model.


\subsection{Downstream task 2: similarity search}
\label{subsec:sim_search}

\begin{figure}
\centering
    \centering
    \includegraphics[width=\columnwidth]{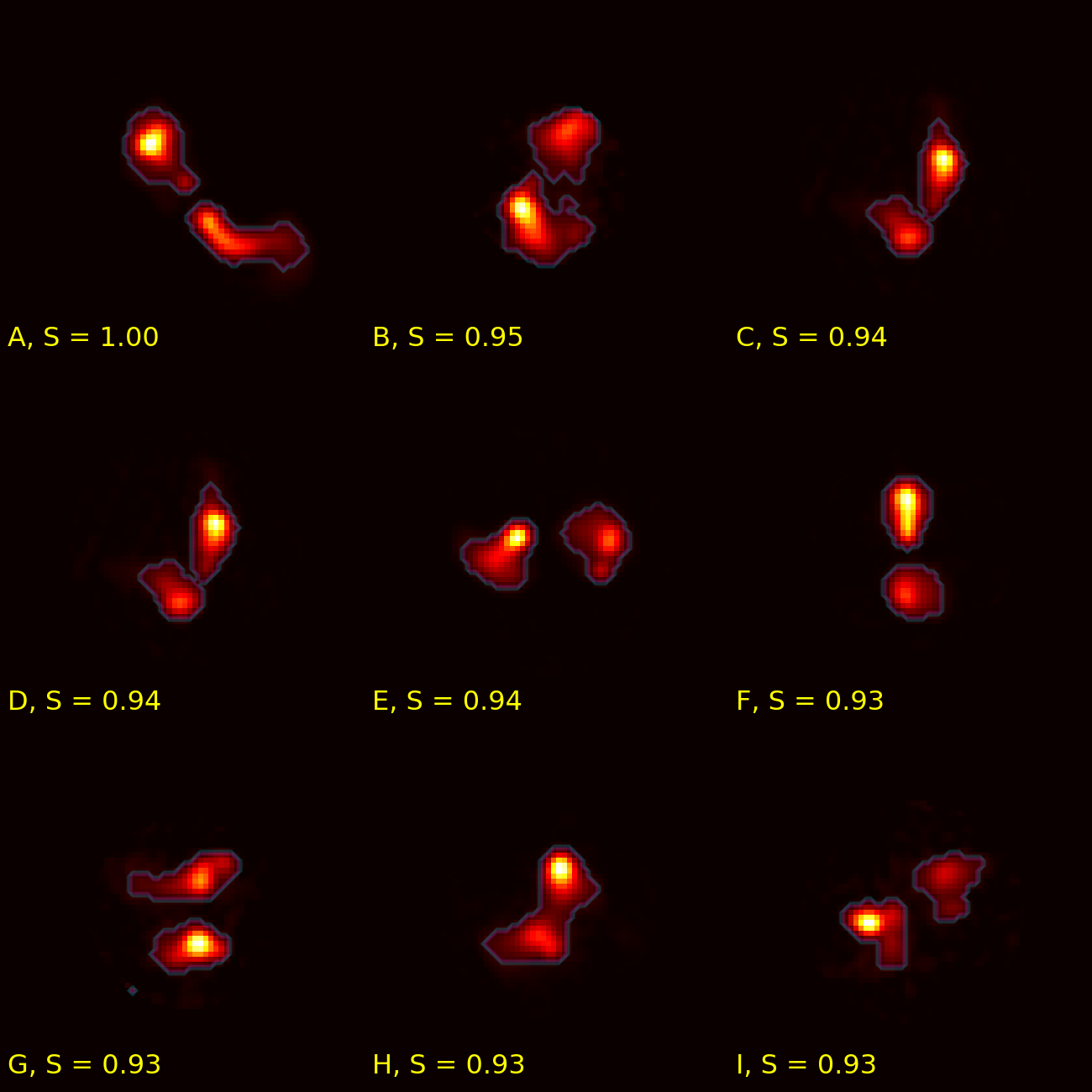}
    \caption{Top left image is the input image into the model. $S$ gives the cosine similarity between the output and input images. We see that our model retrieves hybrid sources with bent tails given an input bent tail hybrid source. Each source has a single bright spot and a single diffuse jet and a bent tail (other than F, which is nonetheless still a similar hybrid source).}
\label{figure:sim_search}
\end{figure}

Drawing inspiration from \citet{Stein2021Self-supervisedDatasets}, we can also take advantage of the semantically meaningful representation from our self-supervised model even if we don't have any labels available for fine-tuning. One possible scientific use case is a similarity search, where we can feed our model a single image of scientific interest and use our representation to extract semantically similar images from the large unlabelled data-set. We can define the "similarity" as the cosine similarity
\begin{equation}
    S^{input, j} = \frac{\vy_\theta^{input} \cdot \vy_\theta^{j}}{|\vy_\theta^{input}| | \vy_\theta^{j} | }
\end{equation}
of the input sample representation $\vy_\theta^{input}$ with the representation, $\vy_\theta^{j}$, of another image (from the unlabelled data). 

To demonstrate that our model is actually solving the task and returning images specific to the input image (rather than just outputting the most common sources in the unlabelled data), we choose a rare hybrid source as our input image. The MiraBest data-set has a ''hybrid`` class which we have no used so far in this work and of which there are only 34 samples. \textit{We do not use a fine-tuned model} - just the weights learned by self-supervised pre-training.  We feed our model with this hybrid image (which has not been seen by the model) which has both FRI and FRII properties, and can be considered out-of-distribution in the supervised case FRI/FRII classification problem. We use the learned representation to return the 8 nearest neighbours in the representation space (shown in Figure~\ref{figure:sim_search}), which we find are not only similar visually, but also have similar physical features, confirming that our representation can be used for finding similar sources. As a bonus, we also find a duplicate data-point in the RGZ DR1 catalogue - even though the data have been (naively) pre-processed to remove duplicates by comparing images pixel wise. This demonstrates that a similarity search could also be used to correct errors where the same object has been included twice due to cataloging differences, e.g. slightly offset cutouts, which are difficult to catch in pre-processing.

To make this visualization of the model's representation space available to radio astronomers, we provide an interactive app where the model can be queried with an arbitrary image from the VLA FIRST survey by inputting sky co-ordinates. We encourage anyone interested in our results to use it - installation takes only a few minutes and no machine learning or coding experience is required to use it. The app can be found on GitHub\footnote{\url{https://github.com/inigoval/rgz-latentexplorer}}.

\subsection{Downstream task 3: cross-survey FR classification}
\label{subsec:cross_survey}

\begin{table*}
\begin{center}
\begin{tabular}{ c|cccc } 
 Method                                         &  \texttt{MIGHTEE}         & \texttt{MiraBest Confident}   & \texttt{MiraBest Uncertain}\\ 
 
 \hline
 
 Supervised \texttt{MIGHTEE}                    & $16.39 \pm 0.16\%$        & $32.64 \pm 0.11\%$            & $34.07 \pm 0.90\%$ \\ 
 
 Supervised \texttt{MiraBest}                   & $14.70 \pm 0.45\%$        & $5.48 \pm 0.35\%$             & $11.37 \pm 0.31\%$\\ 
 
 Pre-train + finetune \texttt{MiraBest}         & $14.00 \pm 0.40 \%$       & $\bm{1.98 \pm 0.32\%}$        & $\bm{8.76 \pm 0.42\%}$\\ 
 
 
Pre-train + finetune \texttt{MIGHTEE}  & $\bm{8.61 \pm 0.11\%}$    & $23.86 \pm 0.78\%$            & $28.84 \pm 0.74\%$ \\

\textbf{Pre-train + finetune \texttt{MIGHTEE + MiraBest}}  & $\bm{8.61 \pm 0.08\%}$    & $4.36 \pm 0.33\%$            & $9.79 \pm 0.29\%$ \\

\end{tabular}
\end{center}
\caption{Test set error rates. We give the training set domain in the Method (left hand) column and the test set domain in the Error (right hand) columns.} 
\label{tab:cross_survey}
\end{table*}

A multi-purpose foundation model should be useful for a wide range of data. To test the ability of the model to generalize well we test the performance of our model on data from a different survey to the RGZ DR1 data-set it was (pre-)trained on. We use a small expert consensus labelled data-set from the MIGHTEE survey (see Section~\ref{subsec:data} for details). It is important to note that these data are observations from a different telescope (MeerKAT; \citet{Jonas2016TheTelescope}) and as such there is some (unquantified) distribution shift expected between the unlabelled and labelled data.

To do this we label our own data-set of FR candidates from the MeerKAT International GHz Tiered Extragalactic Exploration (MIGHTEE) Survey using a consensus of five experts. The sources we label are from the publicly available Continuum Early Science data\footnote{\url{https://www.mighteesurvey.org/data-access}}, details of which can be found in \citet{Jarvis2016TheSurvey, Heywood2022MIGHTEE:Fields}. To obtain our labels, we take the majority classification of the experts and remove point sources and "unsure/too noisy" sources. The final sample contains 72 FRIIs and 45 FRIs, which we use to fine-tune and evaluate our model in Section~\ref{subsec:cross_survey}. We have made this data-set publically available\footnote{\url{https://zenodo.org/record/8188867}}.

Due to the very small size of the labelled data-set, we omit a validation set and simply use the hyper-parameters and settings used when training on the MiraBest data-set (see Section~\ref{subsec:classification}). We use 30\% of the data (35 samples) as a test set (using a smaller proportion yielded noisy results) and the remaining 70\% (92 samples) as a training set. Both of these sets are stratified to contain the same proportion of FRI/FRIIs. We train our model 10 times with a different test/train split and report the average accuracy along with the standard error in Table~\ref{tab:cross_survey} using different combinations of train/test sets. The test/train splits are seeded so they are consistent when testing both the supervised baseline and the fine-tuned foundation model. For comparison, we also provide a baseline where the pre-trained model is fine-tuned on the MiraBest data-set and tested on MIGHTEE. When fine-tuning on the MIGHTEE data-set we train for fewer epochs due to the small amount of labelled data.

We find that pre-training is highly beneficial even in the cross-survey scenario, with the pre-trained models out supervised models from scratchmodels on MIGHTEE test set. When all of the layers are fine-tuned we see an improvement of $\sim8$\% in model accuracy compared with training from scratch, demonstrating that the utility of the pre-trained model is not restricted to data from a single telescope. This large improvement in accuracy highlights the benefits of using our pre-trained model: in the normal supervised case, the data-set is too small to make use of the model's capacity, whereas if we have pre-trained then we can reap the benefits of a larger capacity model without needing extra labelled data or carefully tuning the model hyper-parameters (which would require a validation set, further reducing the already very small training set).

Our results also show that the fine-tuned model transfers better from the MiraBest to the MIGHTEE data-set - this is somewhat unsurprising given that there is close to an order of magnitude more data in the MiraBest data-set. The two best performing models on the MIGHTEE test set are both pre-trained, with one being fine-tuned only on MIGHTEE data and one fine-tuning on both MiraBest and MIGHTEE data. This shows that by fine-tuning on both data-sets the model performs better across both data-sets without losing performance on the smaller MIGHTEE data-set. Although accuracy on the MiraBest test data is lower than when fine-tuned purely on MiraBest data, the overall performance on test sets across both domains is higher.

\subsection{Fine-tuning depth}


\begin{figure}
\centering
    \centering
    \includegraphics[width=\columnwidth]{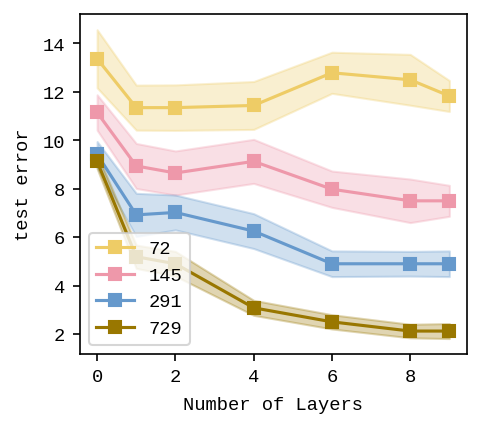}
    \caption{Test set error on the MiraBest \textit{Confident} subset as a function of number of fine-tuned layers. Plotted values are given in \ref{fig:nlayers}. The number of labels used for fine-tuning is given for each curve. Error bars show the standard error}
\label{fig:nlayers}
\end{figure}

When fine-tuning the pre-trained model, we can choose to freeze the deeper layers of the network, omitting them from the fine-tuning. This can be beneficial if we wish to trade some performance for training time, as often fine-tuning the deeper parts of the network does not result in significant performance gain and training is faster when fewer weights are being updated. This effect is particularly pronounced if the scientist training their model does not have access to a GPU and is training their model on a CPU, in which case back-propagation is very computationally expensive and is the dominant term in the training time of the model.

We run experiments to measure test set error when fine-tuning different numbers of layers in our model to inform users of astronomical foundation models when choosing a suitable number of layers to fine-tune for their downstream task. In Figure~\ref{fig:nlayers} we show our results for different label volumes. We show that fine-tuning a single layer can achieve equal performance to the supervised baseline when all weights are trained from scratch and all labels are available. Training with 364 labels on a laptop GPU, fine-tuning a single layer achieves the same test accuracy ($\sim92\%$) but reduces training time by $\sim 66\%$ (from 4min 18s to 1min 29s) compared with training all layers from scratch. This is due to many fewer parameter updates being required and because fewer epochs are needed to train a single layer to convergence compared with an end to end model. However, if a GPU is not available - which can often be the case - the difference in training time is greater. The training time is reduced by $73\%$ from 16m 20s to 4m 22s, which is extremely significant. This increase in difference is because backwards passes for for gradient updates (which are very costly/slow without a GPU) are greatly reduced when fine-tuning a single layer only. In such a case, our foundation model allows a practitioner to leverage large scale models \textit{without needing the compute usually associated with their use}.

However, if compute is not a consideration, Figure~\ref{fig:nlayers} also shows that fine-tuning all layers in the network is beneficial to achieve maximum performance - especially when using many labels. This indicates that our small labeled data-set is large enough to utilise the full capacity of the model. 

Given that some downstream users of our foundation model may wish to use less data than the full MiraBest data-set (e.g. classification of rare objects, difficult to acquire labels), we test the effect of fine-tuning different model depths with varying label volumes also. In this case, we expect the performance of the model to saturate after a certain number of labels are fine-tuned, as the very low amount of labels available simply cannot leverage the full capacity of the model - even when it is only being fine-tuned after pre-training. In Figure~\ref{fig:nlayers}, we see that as the number of labelled data points available for fine-tuning is reduced, model performance improves less as we increase the number of layers fine-tuned. At the lowest label volume (72 labels), we see that there is no benefit in fine-tuning more than one layer.


\subsection{Pre-processing}
\label{subsec:cut_threshold}


\begin{table}
\caption{linear evaluation validation accuracy when pre-training with different cut thresholds. All values are averaged over 10 runs with different seeds for both model parameter initialization and data splitting.}
\label{tab:cut_thresh}
\begin{center}
\begin{tabular}{c|cc} 
 Cut threshold  & \# of training samples     & Error (\%)\\ 
 \hline
 15             & 97513                     & $8.80 \pm 0.56$ \\
 17             & 85860                     & $9.29 \pm 0.60$ \\
 19             & 74482                     & $8.63 \pm 0.59$ \\
 20             & 69015                     & $8.14 \pm 0.43$ \\
 21             & 64071                     & $9.78 \pm 0.59$ \\
 22             & 59668                     & $9.18 \pm 0.58$ \\
 23             & 55559                     & $8.09 \pm 0.35$ \\
 24             & 51910                     & $8.47 \pm 0.50$ \\
 25             & 48774                     & $8.58 \pm 0.68$ \\
 26             & 46044                     & $10.05 \pm 0.59$ \\
 27             & 43417                     & $8.74 \pm 0.53$ \\
 28             & 41007                     & $8.80 \pm 0.52$ \\
\end{tabular}
\end{center}
\end{table}

\begin{figure}
\centering
    \centering
    \includegraphics[width=\columnwidth]{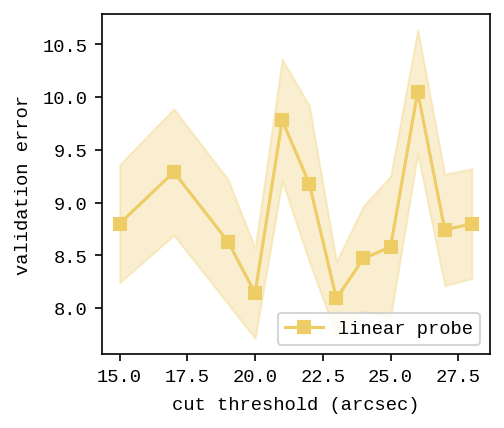}
    \caption{Model performance (validation set error) as a function of the cut threshold used in pre-processing. This is a visualization of Table~\ref{tab:cut_thresh}.}
\label{fig:cut_thresh}
\end{figure}

Even though the RGZ DR1 data-set is unlabelled (with respect to FR classification), a number of physical properties are automatically recorded for each source. One such property is the angular extent of each source, which is given in arcseconds. If the unlabelled data is dominated by low angular extent sources, the quality of the learned representation can suffer \citep{Slijepcevic2022LearningLearning}. The rationale for this is that very small sources are essentially point sources and that including them in the data has the equivalent effect of oversampling a very small sub-set of the available data manifold. In this section, we test the performance of our model when setting different lower limit cut thresholds (removing sources below a specified angular extent).

By doing this, we can remove low extent sources to keep performance high while losing as little available data as possible. It should be noted that this data curation step is performed \textbf{before} the hyper-parameter optimization described in Section~\ref{subsec:augs}, as the optimal hyperparameters will be dependent on the chosen cut threshold.

In Figure~\ref{fig:cut_thresh} (see also Table~\ref{tab:cut_thresh}) we see that exact choice of angular extent cut threshold does not significantly affect the linear separability of our learned representation. Consequently, we arbitrarily choose a lower cut threshold of 20 arcseconds which removes only the very numerous and semantically sparse point sources.


\subsection{Augmentation analysis and hyperparameter optimization}
\label{subsec:augs}
A core part of instance differentiation based pre-training algorithms such as BYOL is the augmentation scheme used to generate the pairs of images. Due to the very different nature of our images compared with standard data-sets such as ImageNet, we use a custom augmentation scheme designed to retain the fine-grained detail of our images while perturbing them enough for the model to learn effectively \citep{Tian2020WhatLearning}. To achieve a more optimal augmentation pipeline and investigate the effect of augmentation pipeline optimization on the resulting learned representation, we tune our augmentation hyper-parameters by searching across multiple possible values and averaging the validation accuracy of the model across multiple initializations of the model/data split.

We only report linear evaluation results as when the model is end-to-end fine-tuned, the small effect of augmentation hyper-parameters on performance is reduced further. This is intuitive as the supervised baseline can achieve high accuracy from randomly initialized parameters. This does however highlight the robustness of the pre-training and fine-tuning paradigm, where the self-supervised foundation model does not need to have optimal linear separation in order to be fine-tuned effectively for a specific downstream task.

\subsubsection{Center cropping}

\begin{figure}
\centering
    \centering
    \includegraphics[width=\columnwidth]{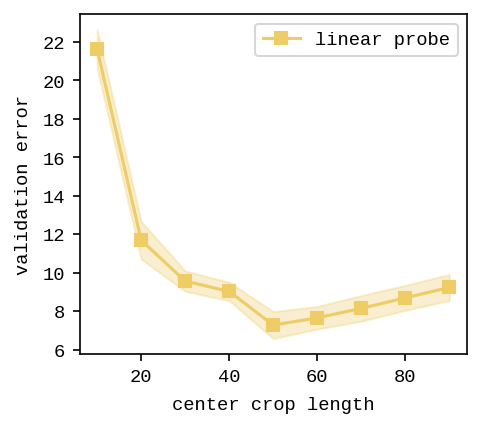}
    \caption{Model performance as a function of the size of each side of the central square of pixels we retain after center cropping the image. Blurring is omitted from the augmentation pipeline for this experiment as it is dependent on our center cropping value.}
\label{fig:center_crop_size}
\end{figure}

The first augmentation in our augmentation pipeline is center cropping the image. The motivation for this is that the raw images are too large - the edges are almost always only zero-valued. By center cropping the image, we can reduce the dimensionality of the input data without reducing the volume of label information, which can be a worthwhile trade-off when training a model. In Figure~\ref{fig:center_crop_size} we see, somewhat surprisingly, that we can achieve moderately high classification performance by using only the central 20x20 pixels of the image (with fine-tuning). This likely reflects the fact that for the majority of sources, information regarding the FRI/FRII classification of the image is concentrated in the center of the image - even if the source extends beyond this. We do however see an increase in performance, as we increase the crop size, before it begins to fall very slowly. Although the maximum performance is achieved at a crop size of 50\,pixels, we use a crop size of 70\,pixels, as the performance is within the uncertainty range of the maximum, but we retain a larger image size. We do this because, in the context of developing a foundation model for various radio astronomy tasks, less cropping might be beneficial for the following reasons: 
\begin{itemize}
    \item Information from extended parts of the source may be beneficial for other tasks which aren't FRI/FRII classification (e.g. similarity search).
    
    \item Rare/complex sources may have more extended features, which will be missed if we choose a small crop size. These kinds of sources do not contribute much to the accuracy metric of the model, but are quite important from a science perspective in the context of detecting unusual sources/structures.
    
\end{itemize}


\subsubsection{Random cropping}

\begin{figure}
\centering
    \centering
    \includegraphics[width=\columnwidth]{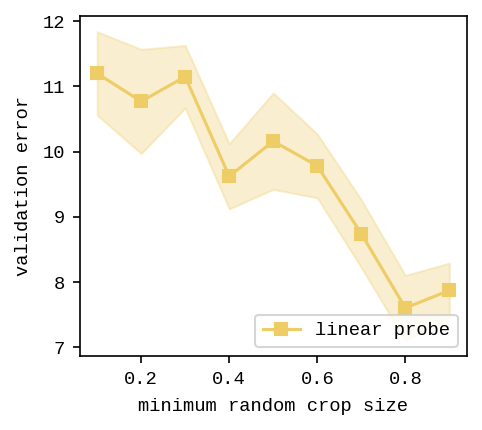}
    \caption{Model performance as a function of the size of each side of the central square of pixels we retain after randomly cropping. Blurring is omitted from the augmentation pipeline for this experiment as it is dependent on our random cropping value.}
\label{fig:random_crop}
\end{figure}

Random cropping is a standard augmentation for contrastive models. It adds stochasticity and increases the difficulty of the instance differentiation task by forcing the model to predict similar representations when provided with only parts of the image. However, these models have generally been optimized in the context of dense terrestrial images where most of the image carries semantic information. In our case, images are sparse and contain many zeros - this can result in meaningless patches of the image being chosen when random cropping which can provide an erroneous learning signal to the model.

We test the performance of our model when varying the minimum possible size of the random crop in our augmentation pipeline. This controls the smallest possible size (as a fraction of image size) that will be chosen when choosing random crops of the image. In Figure~\ref{fig:random_crop} we see that our experimental results validate our intuition regarding image sparsity and aggressive random cropping. We find that using a smaller value for the minimum random crop size has a significant negative effect on the quality of the learned representation. For this reason we opt to use a value of 0.8 for the minimum random crop size - a full order of magnitude higher than the standard value of 0.08 used in the computer science literature \citep{Grill}. 


\subsubsection{Blurring}

\begin{figure}
\centering
    \centering
    \includegraphics[width=\columnwidth]{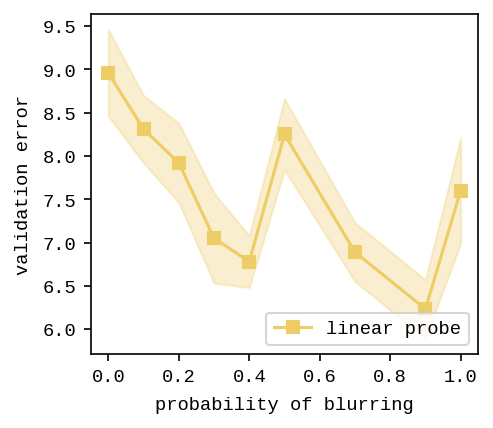}
    \caption{Probability of applying blurring to the augmented images.}
    
\label{fig:blur}
\end{figure}

Blurring is a standard augmentation for contrastive models to add difficulty to the instance differentiation task. However, due to the very fine-grained nature of astronomical images, blurring can destroy a significant portion of the information in the image and in some extreme cases cause two different images to look quite similar after they have been blurred. We test whether blurring improves linear evaluation on the FR classification pipeline by varying the probability of blurring during data augmentation and recording validation accuracy.

 
In Figure~\ref{fig:blur} we show that varying the blurring parameter has an effect on the resulting representation. Although there is a trend towards using a larger blurring probability, our results are too stochastic to be certain. Furthermore, the difference in performance is relatively small.


Intuitively we expect that difficult to classify sources or more fine-grained classification may require more fine-grained detail that can be lost with aggressive blurring. This effect may not be detected when performing FR classification due to the coarseness of the classes. For this reason we choose a low blurring parameter as more difficult downstream tasks (such as sub-class classification) may require more fine-grained detail.


\subsection{Model Size}

\begin{table}
\caption{linear evaluation classification results with our BYOL model when using different sized models. All models use the ResNet architecture, just with a differing number of layers.}
\label{tab:modelsize}
\begin{center}
\begin{tabular}{ c|cc } 
 Layers & Error  \\ 
 \hline
 14     & $9.58 \pm 0.64\%$ \\
 18     & $8.20 \pm 0.64\%$ \\ 
 34     & $8.20 \pm 0.57\%$ \\ 
\end{tabular}
\end{center}
\end{table}

In computer vision, improvements are often made by simply scaling models (increasing width/depth). However, in astronomy our images are sparse in the pixel space, comparatively low resolution and often have only a single color channel, all resulting in less semantically dense images. For this reason, we test whether increasing model size actually improves our model performance, as increasing the number of parameters increases the cost of training the models and can lead to overfitting. We hope that our experiments will guide future work to allow faster iteration through choosing the smallest possible high performance model.

Table~\ref{tab:modelsize} shows that increasing the model size for our data-set has minimal benefits, with quickly diminishing returns as we increase the number of layers above the ResNet-18 model. For this reason, we use ResNet-18 as a our architecture for the remainder of our experiments, and suggest that practitioners with similar radio astronomical data-sets do the same.


\subsection{Augmentation ablation study}
\begin{table*}
\begin{center}
\begin{tabular}{ccccc|cc} 
 Random Rotation    & Random Resized Crop   & Random Flipping   & Color Jitter  & Blur & Linear Evaluation Error (\%)    & Fine-tuning Error (\%) \\ 
 \hline
 
\vspace{2mm}

\checkmark         &   \checkmark          & \checkmark        & \checkmark   & \checkmark & $8.33 \pm 0.40$  & $ 2.22 \pm 0.31 $  \\ 

\texttimes         &   \checkmark          & \checkmark        & \checkmark    & \checkmark & $6.44 \pm 0.64$  &  $2.31 \pm 0.21$ \\ 

\checkmark         &   \texttimes          & \checkmark        & \checkmark    & \checkmark & $13.46 \pm 0.64$  &  $4.33 \pm 0.33$ \\ 

\checkmark         &   \checkmark          & \texttimes        & \checkmark    & \checkmark & $7.40 \pm 0.64$  &  $2.31 \pm 0.41$ \\ 

\checkmark         &   \checkmark          & \checkmark        & \texttimes    & \checkmark & $8.37 \pm 0.35$  &  $3.17 \pm 0.29$ \\ 

\vspace{2mm}

\checkmark         &   \checkmark          & \checkmark        & \checkmark    & \texttimes & $8.08 \pm 0.60$  & $3.28 \pm 0.38$ \\ 

\checkmark         &   \checkmark          & \texttimes        & \texttimes    & \checkmark & $9.81 \pm 0.28$  & $3.75 \pm 0.30$\\ 

\checkmark         &   \texttimes          & \checkmark        & \texttimes    & \checkmark & $12.86 \pm 0.30$  & $3.75 \pm 0.70$ \\ 

\vspace{2mm}

\texttimes         &   \checkmark          & \checkmark        & \texttimes    & \checkmark & $6.63 \pm 0.64$  &  $3.75 \pm 0.30$\\ 

\checkmark         &   \texttimes          & \texttimes        & \texttimes    & \texttimes & $8.85 \pm 0.51$  & $5.38 \pm 0.41$ \\ 

\texttimes         &   \checkmark          & \texttimes        & \texttimes    & \texttimes & $8.94 \pm 0.25$  &  $5.10 \pm 0.25 $ \\ 
 
\texttimes       &   \texttimes          & \checkmark        & \texttimes    & \texttimes & $23.75 \pm 0.35$  & $ 10.00 \pm 0.29 $  \\ 

\texttimes         &   \texttimes          & \texttimes        & \checkmark    & \texttimes & $18.06 \pm 0.68$  &  $6.06 \pm 0.38$ \\ 

\texttimes         &   \texttimes          & \texttimes        & \texttimes    & \checkmark & $20.96 \pm 0.81$  &  $3.85 \pm 0.35 $ \\ 

\vspace{2mm}

\texttimes         &   \texttimes          & \texttimes        & \texttimes    & \texttimes & $24.52 \pm 0.58$  &  $6.35 \pm 0.33 $ \\ 


\end{tabular}
\end{center}
\caption{An ablation study of augmentations using test-set accuracy trained with all available training labels (no validation set).}
\label{tab:augablate}
\end{table*}

In order to identify which of the augmentations described in Section~\ref{subsec:augs} are most crucial to learning, and which can destroy too much information in the images, we ablate them i.e. we remove individual augmentations and measure the performance of the resulting model. The results of our ablation studies are presented in Table~\ref{tab:augablate}.

We find that while all of the augmentations contribute to some degree to the model's performance, random resized cropping is has the most significant effect and note that we are able to learn a good representation with random resized crop as our only augmentation. Random rotation, while having less effect when ablated out, also allows the model to learn a good representation when included as the only augmentation. 

We also see that the differences in linear evaluation accuracy from ablation in the augmentation scheme become much smaller when we fine-tune the model. It may seem counter intuitive that a model based on instance differentiation with no augmentations at all can achieve over $90 \% $ accuracy when fine-tuned, but this needs to be contextualized by the fact that a supervised model is able to achieve better performance from randomly initialized parameters from scratch. As a worst case scenario with no augmentations at all, pre-training reverts to a simple student-teacher model with fine-tuning in a fully supervised way albeit with a non-optimal layer wise decayed learning rate. This highlights the robustness of pre-training, showing that the model will perform similarly to the baseline case even if the pre-training step is sub-optimal.

Overall we find that while we can obtain small gains in validation accuracy from choosing optimal hyper-parameters, model performance is significantly less sensitive to hyper-parameters (or even complete omission of augmentations) when the model is end-to-end fine-tuned compared to when the representation is probed using linear evaluation. The fine-tuned model is generally able to learn a good representation as long as random rotation and random cropping are not removed. This bodes well for applications of contrastive learning in the wider science community, as it shows that reasonably chosen hyper-parameters can achieve very good performance when the model is fine-tuned and augmentations do not need to be painstakingly designed for every new data-set.


\section{Conclusion}
We show that given a downstream task, our foundation model out-performs the baseline at all label volumes. With all labels available, the end-to-end fine-tuned model solves the MiraBest Confident FRI/FRII classification problem efficiently, achieving a test set error of $1.92 \pm 0.25$ on the held out test set, $\sim 3 \%$ less than the baseline which more than halves the baseline test error, whil also generalizing better to \textit{Uncertain} data. This is a very significant margin of improvement given the inherent label noise in the data-set, which may set a hard limit on the highest possible performance. 

Our foundation model also allows the practitioner to choose trade-offs suitable for their application. For example, if compute is plentiful but human labels are scarce, an end-to-end fine-tuned foundation model significantly outperforms the supervised baseline (regardless of how many labels are available). In another case, researchers may want to test a new classification scheme but may not have access to a GPU for training large models - in this case a single layer of our foundation model can be fine-tuned on a laptop to achieve performance comparable to a large network trained from scratch. We believe that the ability to easily customise the training scheme of the model to achieve maximal performance given the resources available is an extremely powerful characteristic of building foundation models in this way and an important step towards making data analysis of large scientific data-sets accessible to a larger number of practitioners.

We also provide an example scientific use case for our model, which only requires a single input data point of interest. Given a rare source, our model is able to retrieve multiple samples with similar physical features. We emphasise that this functionality is computationally very cheap (requires no training or GPU) and that we have provided a front end enable minimal friction for those who wish to use it. 

We further test the generalizability of our model by fine-tuning and evaluating on label scarce FR classification data-set from a different telescope/survey (MeerKAT/MIGHTEE): finding that our model yields a $\sim8\%$ improvement over the baseline on a label scarce FR classification task. This demonstrates that our model's usefulness is not limited to a single telescope or survey and can be straightforwardly fine-tuned to data from different sources.

Overall we have demonstrated that the learned representation is applicable to a wide variety of tasks, and we hope that the astronomy community will build upon our task-agnostic model with new tasks/applications beyond what we have tested.

We show that while it is important to choose a set of suitable augmentations with reasonable parameters - especially if we are only using a linear evaluation layer or fine-tuning a small number of layers - an end-to-end fine-tuned model is not overly sensitive to hyper-parameter choices. Model performance only drops below the supervised base-line performance when all but one augmentation are completely removed, which likely makes the instance differentiation task too easy to solve without learning meaningful features. This justifies the use of our foundation model for a wide variety of tasks without the possibility of a sub-optimal pre-training stage negatively affecting performance - which can be an issue with single stage semi-supervised methods \citep{Slijepcevic2022RadioShift}.

We provide visualizations of the RGZ DR1 and MiraBest data-sets, showing that the MiraBest data-set is only a small part of the full data manifold, and that exposing our model to the full RGZ DR1 data improves both in-distribution and out-of-distribution classification performance after fine-tuning. We also show that the representation space learned during pre-training is structured with respect to source extension, showing that the model has learned valuable semantic information by solving the instance differentiation task during pre-training. We use this structure to find multiple hybrid sources in the RGZ DR1 data-set with only a single hybrid source as input.

In summary, we train a self-supervised model on a large unlabelled data-set of sources from the VLA FIRST survey as a first step towards building a foundational model for radio astronomy. We demonstrate that the model measurably improves model performance on classification tasks even when fine-tuning on data from a different source, and is a powerful tool for one-shot similarity search. Furthermore, the pre-training in this way allows flexibility with regards to trade offs in compute, labelling cost and model performance.

In future work, we would like to integrate data from a wider range of sources/domains which could include different modalities (e.g. spectral information). This could enable the model to be applied to a greater range of tasks and to learn relationships between different data.



\section*{Acknowledgements}

IVS, AMS, MB \& MW gratefully acknowledge support from the UK Alan Turing Institute under grant reference EP/V030302/1. IVS gratefully acknowledges support from the Frankopan Foundation.

This work has been made possible by the participation of more than 12,000 volunteers in the Radio Galaxy Zoo Project. The data in this paper are the result of the efforts of the Radio Galaxy Zoo volunteers, without whom none of this work would be possible. Their efforts are individually acknowledged at \url{ http://rgzauthors.galaxyzoo.org}.

\section*{Data Availability}
Code for this paper can be found at \url{https://github.com/inigoval/}. Model weights for use in other work can be found at \url{https://zenodo.org/record/7615104}.

The RGZ DR1 catalogue will be made publicly-available through Wong et al (2023; in preparation). This work makes use of the MiraBest machine learning dataset, which is publically available under a Creative Commons 4.0 license at \url{https://doi.org/10.5281/zenodo.4288837}.

The labelled data set created from the MIGHTEE Early Science Data can be found at \url{https://zenodo.org/record/8188867}.



\bibliographystyle{rasti}
\bibliography{references} 




\appendix


\bsp	
\label{lastpage}
\end{document}